\documentclass{article}
\usepackage[utf8]{inputenc}
\usepackage[a4paper,left=3cm,right=3cm,top=3cm,bottom=3cm]{geometry}
\usepackage{amsmath}
\usepackage{enumitem}
\usepackage{tikz}
\usepackage{cite}

\usepackage{floatrow}
\usepackage{caption}

\usepackage{amssymb}
\usepackage{amsbsy}
\usepackage{bm}
\usepackage{fixmath}
\usepackage{graphicx}
\usepackage{hyperref}
\usepackage{booktabs}
\usepackage{xspace}

\newcommand{\be}{\begin{equation}}
\newcommand{\ee}{\end{equation}}

\definecolor{red}{rgb}{1,0,0}

\def\bea{\begin{eqnarray}}
\def\eea{\end{eqnarray}}
\definecolor{kkcolor}{rgb}{1,0,0}

\newcommand\kkout{\marginpar{\color{kkcolor}$\int$}\bgroup\markoverwith{\color{kkcolor}{\rule[0.4ex]{2pt}{0.8pt}}}\ULon}

\definecolor{pkcolor}{rgb}{0,0,1}

\newcommand\pkout{\marginpar{\color{pkcolor}$\int$}\bgroup\markoverwith{\color{pkcolor}{\rule[0.4ex]{2pt}{0.8pt}}}\ULon}

\definecolor{hkcolor}{rgb}{0.7,0.0,0.0}

\newcommand\hkout{\marginpar{\color{hkcolor}$\clubsuit$}\bgroup\markoverwith{\color{hkcolor}{\rule[0.4ex]{2pt}{0.8pt}}}\ULon}

\definecolor{avhcolor}{rgb}{0.7,0.7,0.0}

\newcommand\avhout{\marginpar{\color{avhcolor}$\clubsuit$}\bgroup\markoverwith{\color{avhcolor}{\rule[0.4ex]{2pt}{0.8pt}}}\ULon}

\hypersetup{
  colorlinks   = true,    
  urlcolor     = blue,    
  linkcolor    = blue,    
  citecolor    = black     
}

\title{The Balitsky-Kovchegov equation for dipole gluon density in the momentum space}
\author{Krzysztof Kutak\\
{\it Institute of Nuclear Physics, Polish Academy of Sciences} \\
     {\it  Radzikowskiego 152, 31-342 Krak\'ow, Poland } \\ \\
}
\date{January 2026}

\begin{document}

\maketitle
\begin{abstract}
I present derivation the BK equation for the dipole gluon density in momentum space, starting from its standard formulation in coordinate space. I review the equation for both proton and nuclear targets, and I also discuss the resummed BK evolution.The purpose of this paper is to consolidate derivations and formulas scattered across the literature, to show in detail how the structure of the triple-Pomeron vertex emerges in the unfolded form of the nonlinear term, and to establish a consistent notation throughout.
\end{abstract}

\section*{Introduction}
Deep inelastic scattering experiments at HERA determined that, as Bjorken 
x decreases, parton densities grow steeply \cite{H1:1995f2}. This was consistent with predictions based on the Balitsky, Fadin, Kuraev, Lipatov (BFKL) equation \cite{Balitsky:1978ic,Kuraev:1977fs}, which indeed predicts such growth. However, soon after the BFKL equation was proposed, it was argued that an unlimited growth of parton densities would lead to a violation of unitarity and that partons have to saturate\cite{Gribov:1983ivg,Mueller:1985wy} (for review see \cite{Kovchegov:2012mbw}). Following this line of thought, new equations were proposed to generalize BFKL by including a nonlinear term \cite{Gribov:1983ivg}. This early proposal was later refined with the formulation of the Color Glass Condensate effective theory (for a review see \cite{Gelis:2010nm} and references therein), whose basic evolution equations are the Balitsky, Kovchegov (BK) equation \cite{Balitsky:1995ub,Kovchegov:1999ua} and the Jalilian-Marian,Iancu, McLerran, Weigert, Leonidov, Kovner (JIMWLK) equation \cite{Jalilian-Marian:1997qno,Ferreiro:2001qy}.

In particular, the BK equation was formulated as an evolution equation for the dipole amplitude, which describes the fluctuation of a virtual photon into a quark–antiquark pair and its evolution with increasing energy. It is formulated in mixed space, where the dipole amplitude depends on rapidity and on the transverse positions of the quark and antiquark. This formulation has clear advantages: the kernel is the same for the linear and nonlinear parts, the equation has fixed points at zero and one, and next-to-leading-order corrections \cite{Balitsky:2007feb}—as well as, more recently, resummations with an appropriate choice of basis—have been computed to stabilize the kernel and account for higher-order effects \cite{Boussarie:2025bpq}. However, the momentum-space formulation is useful as well. In momentum space, one can easily see how the nonlinear term prevents diffusion into the infrared \cite{Golec-Biernat:2001dqn} and why the gluon density is suppressed as the transverse momentum becomes small. Moreover, the momentum-space formulation provides direct access to gluon kinematics and allows for an unambiguous implementation of kinematical corrections in the emission part of the kernel \cite{Kwiecinski:1997ee,Li:2022avs,Kutak:2003bd}. In addition, DGLAP corrections can be implemented conveniently.

The paper is organized as follows. In Sec.~1, I present the BK equation in mixed space and the BK equation in momentum space with a local nonlinear term. Next, I present in detail the transformation leading to an evolution equation for the dipole gluon density, both for a proton and for a nuclear target. In Sec.~2, I present the BK equation in resummed form. A summary is given in Sec.~3.
\section{From BK in position space to momentum space}
The Balitsky-Kovchegov evolution equation for the dipole amplitude \cite{Balitsky:1995ub,Kovchegov:1999ua} reads
\begin{multline}
    \frac{\partial}{\partial \ln 1/x} N(\vec{x}_T,\vec{y}_T;x) =\frac{\alpha_s N_c}{2\pi} \int\! d^2z_T\,  \frac{(\vec{x}_T-\vec{y}_T)^2}{(\vec{x}_T-\vec{z}_T)^2(\vec{y}_T-\vec{z}_T)^2} \,  \Bigg\{ N(\vec{x}_T,\vec{z}_T;x) + N(\vec{y}_T,\vec{z}_T;x)\\ - N(\vec{x}_T,\vec{y}_T;x) - N(\vec{x}_T,\vec{z}_T;\eta)N(\vec{z}_T,\vec{y}_T;x) \Bigg\} \,.
    \label{eq:BK_equation}
\end{multline}
The kernel of the equation accounts for splitting of initial dipole which ends are at positions $\vec x_T$ and $\vec y_T$ into two daughter dipoles which ends are located at positions $\vec x_T$, $\vec z_T$  $\vec y_T$ and $\vec z_T$.  The nonlinear term prevents for unconstrained growth of the dipole amplitude as energy increases and limits it to maximal value equal one. The equation has been extensively studied numerically in \cite{Golec-Biernat:2003naj,Berger:2010sh,Cepila:2024qge,Lappi:2015fma}\\
From the position space BK equation one can obtain equation for transverse momentum dependent dipole gluon density.
This  quantity is basic object which enters factorization formula for DIS reduced cross section as well is a quantity from which various TMDs can be calculated that enter Improved Transverse Momentum Dependent factorization \cite{Kotko:2015ura,vanHameren:2016ftb,Altinoluk:2019fui} as well as small-$x$ Transverse Momentum Dependent factorization \cite{Dominguez:2011wm}. 
The steps outlined below will closely follow 
 the results obtained in \cite{Kutak:2003bd,Bartels:2007dm}.
 We will use $k^2=|\vec{k}_T|^2$ as an argument of gluon density. This is the standard notation used in the discussion of angular averaged distributions.\\
Let's define the following Fourier transforms \cite{Kimber:2001nm,Kutak:2003bd}
\begin{equation}
{\cal F}(x,k^2)=\frac{N_c}{\alpha_s (2\pi)^3}\int d^2b\int d^2re^{ik\cdot r}\nabla^2_{r}\,N(r,b,x)
\end{equation}
\begin{equation}
\Phi(x,k^2)=\frac{1}{2\pi}\int d^2b \int\frac{d^2 r}{r^2}e^{i k \cdot r}N(r,b,x)
\label{eq:dipolampl2}
\end{equation}
where 
$ \vec{r}_T=(\vec{x}_T-\vec{y}_T)/2$ and $
    \vec{b}_T=(\vec{x}_T+\vec{y}_T)/2 \,.$
The function ${\cal F}(x,k^2)$ is the dipole gluon density.
Here the function $\Phi(x,k^2)$ plays a role of auxiliary quantity 
for which the BK equation has a simple form with local nonlinear term as will be shown below. 
The explicit relation between the two functions is
${\cal F}(x,k^2)$ and $\Phi(x,k^2)$
is
\begin{equation}
{\cal F}(x,k^2)=\frac{N_c}{4\alpha_s \pi^2}k^2\nabla^2_{k}\Phi(x,k^2)
\end{equation}
and it's inverse
\begin{equation}
\Phi(x,k^2)=\frac{\alpha_s\pi^2}{N_c}\int_{k^2}^{\infty}\frac{dl^2}{l^2}\ln\frac{l^2}{k^2}{\cal F}(x,l^2).
\end{equation}
It has been shown in \cite{Kovchegov:1999ua} that once one considers homogeneous infinite nucleus the $b$ integral in the eq.  (\ref{eq:BK_equation}) enters only via initial condition and one has 
\begin{equation}
    \Phi(x,k^2)=\frac{1}{2\pi}\int d^2b \int\frac{d^2 r}{r^2}e^{i k \cdot r}N(r,b,x)=\frac{1}{2\pi}\int d^2b \int\frac{d^2 r}{r^2}e^{i k \cdot r}N(r,x)=
   \int d^2b\,\Phi_b(x,k^2)
   \label{eq:bdep}
\end{equation}
the function $\Phi_b(x,k^2)$ obeys the following equation
\begin{equation}
\begin{split}
\frac{\partial \Phi_b(x,k^2)}{\partial\ln 1/x}= \overline\alpha_s
\int_0^{\infty}\frac{dl^2}{l^2}
\bigg[\frac{l^2\Phi_b(x,l^2)- k^2\Phi_b(x,k^2)}{|k^2-l^2|}+ \frac{
k^2\Phi_b(x,k^2)}{\sqrt{(4l^4+k^4)}}\bigg]
-\overline\alpha_s\Phi^2_b(x,k^2).
\label{eq:wwglue}
\end{split}
\end{equation}
where $\overline\alpha=N_c\alpha_s/\pi$, which is obtained by Fourier transform of eq. (\ref{eq:dipolampl2}) \cite{Kovchegov:1999ua,Marquet:2005zf}
\footnote{Detailed presentation of the Fourier transform relating eq. (\ref{eq:BK_equation}) and eq. (\ref{eq:wwglue}) can be found in H. Mantysaari master thesis  {\it Balitzky-Kovchegov equation} \url{https://jyx.jyu.fi/jyx/Record/jyx_123456789_37095}}.\\
The equation for ${\cal F}(x,k^2)$
is obtained from eq. (\ref{eq:wwglue}) in several steps. First inserting in the nonlinear part of it relation expressing $\Phi_b (x,k^2)$  in terms of ${\cal F}_b(x,k^2)$ (note that we use relation between quantities unintegrated over impact parameter) and acting on the whole equation with the operator \cite{Kutak:2012qk}:

\begin{equation}
\frac{N_{c}}{4\,\alpha_{s}\,\pi^{2}}\,
k^{2}\,\nabla^{2}_{\!k}\,\Phi_b(x,k^{2}) \, .
\end{equation}
that transforms $\Phi_b(x,k^2)$ to  ${\cal F}_b(x,k^2)$. For method based on Mellin transform see \cite{Kutak:2006rn}.

\noindent
The linear part of the equation can be written in the form of a differential operator. We have:
\begin{equation}
\frac{\partial\Phi_b(x,k^{2})}{\partial\ln1/x}
=
\alpha_{s}\,
\chi\!\left(-\frac{\partial}{\partial\ln k^{2}}\right)\,
\Phi_b\!\left(x,k^{2}\right)
-\bar\alpha_s,
\Phi^{2}_b\!\left(x,k^{2}\right),
\label{eq:BK-diffop}
\end{equation}
where
\begin{equation}
\chi(\gamma)=2\psi(1)-\psi(\gamma)-\psi(1-\gamma)
\label{eq:chi-def}
\end{equation}
is the characteristic function of the BFKL kernel.
We represent the BFKL kernel of the BK equation as a power series around $\gamma_{c}=0.373$:
\begin{align}
\chi\!\left(-\partial_{\ln k^{2}}\right)\Phi_b(x,k^{2})
&=
\Bigg[
\chi(\gamma_{c})
+\bigl(-\partial_{\ln k^{2}}-\gamma_{c}\bigr)\chi'(\gamma_{c})
+\frac{1}{2!}\bigl(-\partial_{\ln k^{2}}-\gamma_{c}\bigr)^{2}\chi''(\gamma_{c})
+\cdots
\Bigg]\Phi_b(x,k^{2}),
\label{eq:chi-expansion}
\end{align}

The laplacian can be pulled through the derivative and acts directly on $\Phi_b$. After that the full kernel can be restored. The nonlinear term requires explicit evaluation of laplacian acting on the integral operator.


\begin{equation}
\begin{aligned}
\frac{\partial \mathcal{F}_b(x,k^{2})}{\partial \ln(1/x)}
&=
\frac{N_c\alpha_s}{\pi}\int_{0}^{\infty}\frac{dl^{2}}{l^{2}}
\Bigg[
\frac{l^{2}\mathcal{F}_b(x,l^{2})-k^{2}\mathcal{F}_b(x,k^{2})}{\lvert k^{2}-l^{2}\rvert}
+\frac{k^{2}\mathcal{F}_b(x,k^{2})}{\sqrt{4l^{4}+k^{4}}}
\Bigg]
\\[0.6ex]
&\quad
-\frac{1}{4}\pi\,\alpha_{s}^{2}\,k^{2}\,
\nabla^{2}_{\!k}
\left[
\left(
\int_{k^{2}}^{\infty}\frac{dl^{2}}{l^{2}}\,
\ln\!\frac{l^{2}}{k^{2}}\,
\mathcal{F}_b(x,l^{2})
\right)^{2}
\right].
\label{eq:funb}
\end{aligned}
\end{equation}
The BK equation in this form has been for the first time presented in 
\cite{Bartels:2006ea,Kutak:2006rn}. For another derivation see \cite{Nikolaev:2006za}.
\begin{figure}[t]
  \centering
  \begin{minipage}[t]{0.3\textwidth}
    \centering
    \includegraphics[width=\linewidth]{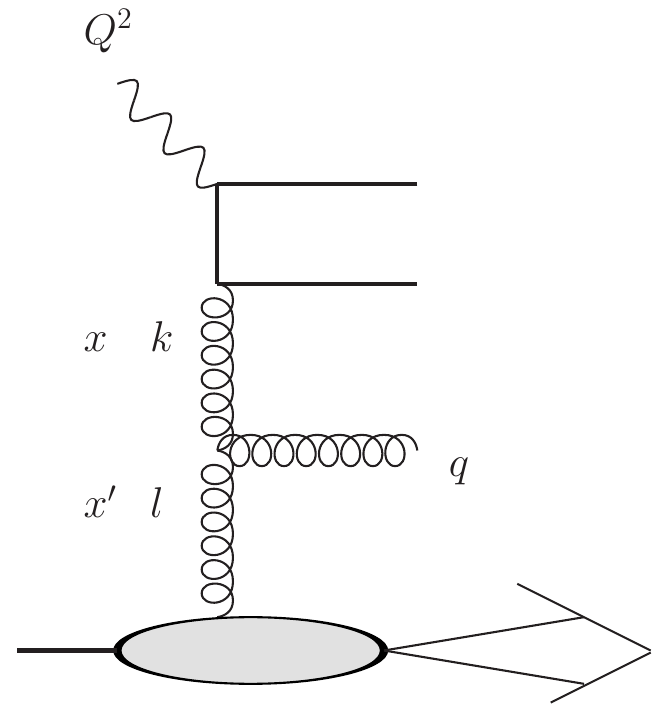}
  \end{minipage}%
  \hspace{15mm}
  \begin{minipage}[t]{0.25\textwidth}
    \centering
    \includegraphics[width=\linewidth]{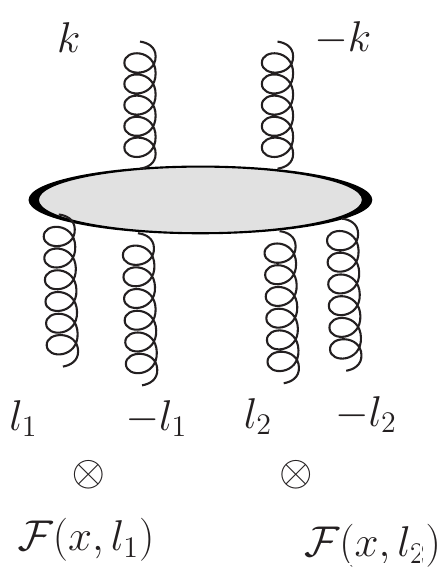}
  \end{minipage}
  \caption{Left: Kinematical variables. Right: Triple pomeron vertex convoluted with gluon
density.}
  \label{fig:plot1-tpv}
\end{figure}
While the equation does not depend dynamically on the impact parameter one needs to integrate over in order to calculate cross sections for quantities which are impact parameter independent (one can see this already in eq. (\ref{eq:bdep}). In the phenomenology based on rcBK equation \cite{Albacete:2010sy} this is usually done by 
\begin{equation}
    \int d^2b\rightarrow S_\perp
\end{equation}
where $S_\perp$ is the transverse area of the proton or nucleus. 
However, in \cite{Kutak:2003bd} another approach has been developed.
The goal was to obtain equation for quantity that is normalized by transverse area.
For cylinder like target, one may use the following ansatz for factorization \cite{Kutak:2003bd}
\begin{equation}
    {\cal F}_b(x,k^2)={\cal F}(x,k^2)S(b)
    \label{eq:fact}
\end{equation}
with normalization conditions
\begin{equation}
\int d^2{\bf b}\,S(b)=1,\,\,\,\int d^2{\bf b}\,S^2(b)=\frac{1}{\pi R_p^2}
\label{eq:norm}
\end{equation}
where $S(b)$ is the profile  function $ S(b)=\theta(R-b)/\pi R_p^2$ with $R_p$ radius of proton.\footnote{ 
This choice is of course not unique. In the \cite{Kutak:2006rn} Gaussian profile was also considered}\\
Once the factorized formula (\ref{eq:fact}) is substituted into eq. (\ref{eq:funb}) and the normalization conditions (\ref{eq:norm}) are used we obtain

\begin{equation}
\frac{\partial \mathcal{F}(x,k^{2})}{\partial \ln(1/x)}
=\frac{N_c\alpha_{s}}{\pi}\int_{0}^{\infty}\frac{dl^{2}}{l^{2}}
\left[\frac{l^{2}\mathcal{F}(x,l^{2})-k^{2}\mathcal{F}(x,k^{2})}{\lvert k^{2}-l^{2}\rvert}
+\frac{k^{2}\mathcal{F}(x,k^{2})}{\sqrt{4l^{4}+k^{4}}}\right]
-\frac{\alpha_{s}^{2}k^{2}}{4R_p^{2}}\,\nabla^{2}_{\!k}\left(\int_{k^{2}}^{\infty}\frac{dl^{2}}{l^{2}}\ln\!\frac{l^{2}}{k^{2}}\,\mathcal{F}(x,l^{2})\right)^{2}.
\label{eq:BKmomentum}
\end{equation}
In the above $\nabla_{k^2}^2=4\left(\frac{\partial }{\partial k^2}+k^2\frac{\partial^2}{\partial( k^2)^2}\right)$.
Let's calculate the first and second derivative of the nonlinear term in eq. (\ref{eq:BKmomentum})
\begin{equation}
\begin{aligned}
\frac{\partial}{\partial k^{2}}
\left(
\int_{k^{2}}^{\infty}\frac{dl^{2}}{l^{2}}\,
\ln\!\frac{l^{2}}{k^{2}}\,
\mathcal{F}(x,l^{2})
\right)^{2}
&=
-\frac{2}{k^{2}}
\left(
\int_{k^{2}}^{\infty}\frac{dl^{2}}{l^{2}}\,
\ln\!\frac{l^{2}}{k^{2}}\,
\mathcal{F}(x,l^{2})
\right)
\left(
\int_{k^{2}}^{\infty}\frac{dl^{2}}{l^{2}}\,
\mathcal{F}(x,l^{2})
\right).
\end{aligned}
\end{equation}

\begin{equation}
\begin{aligned}
\frac{\partial^{2}}{\partial (k^{2})^{2}}
\left(
\int_{k^{2}}^{\infty}\frac{dl^{2}}{l^{2}}\,
\ln\!\frac{l^{2}}{k^{2}}\,
\mathcal{F}(x,l^{2})
\right)^{2}
&=
2\Bigg[
\left(
\frac{1}{k^{2}}
\int_{k^{2}}^{\infty}\frac{dl^{2}}{l^{2}}\,
\mathcal{F}(x,l^{2})
\right)^{2}
\\[0.6ex]
&\qquad\quad
+
\frac{1}{(k^{2})^{2}}
\left(
\int_{k^{2}}^{\infty}\frac{dl^{2}}{l^{2}}\,
\ln\!\frac{l^{2}}{k^{2}}\,
\mathcal{F}(x,l^{2})
\right)
\left(
\mathcal{F}(x,k^{2})
+
\int_{k^{2}}^{\infty}\frac{dl^{2}}{l^{2}}\,
\mathcal{F}(x,l^{2})
\right)
\Bigg].
\end{aligned}
\end{equation}
Combining the second order and first order derivatives together we get
\begin{equation}
\begin{aligned}
\label{eq:nabla}
\nabla^{2}_{\!k^2}
\left[
\left(
\int_{k^{2}}^{\infty}\frac{dl^{2}}{l^{2}}\,
\ln\!\frac{l^{2}}{k^{2}}\,
\mathcal{F}(x,l^{2})
\right)^{2}
\right]
&=
\frac{8}{k^{2}}
\Bigg\{
\left(
\int_{k^{2}}^{\infty}\frac{dl^{2}}{l^{2}}\,
\mathcal{F}(x,l^{2})
\right)^{2}
\\[0.6ex]
&\qquad\quad
+\;
\mathcal{F}(x,k^{2})
\left(
\int_{k^{2}}^{\infty}\frac{dl^{2}}{l^{2}}\,
\ln\!\frac{l^{2}}{k^{2}}\,
\mathcal{F}(x,l^{2})
\right)
\Bigg\}.
\end{aligned}
\end{equation}
Inserting eq. (\ref{eq:nabla}) in the in the eq.  (\ref{eq:BKmomentum}) we obtain \cite{Kutak:2003bd,Bartels:2007dm}
\begin{equation}
\label{eq:faneq1}
\begin{aligned}
\frac{\partial \mathcal{F}(x,k^{2})}{\partial \ln(1/x)}
&=
\frac{N_c\alpha_s}{\pi}\int_{0}^{\infty}\frac{dl^{2}}{l^{2}}
\left[
\frac{l^{2}\mathcal{F}(x,l^{2})-k^{2}\mathcal{F}(x,k^{2})}{\lvert k^{2}-l^{2}\rvert}
+\frac{k^{2}\mathcal{F}(x,k^{2})}{\sqrt{4l^{4}+k^{4}}}
\right]
\\[0.6ex]
&\quad
-\frac{2\,\alpha_{s}^{2}}{R_p^{2}}
\Bigg\{
\left(\int_{k^{2}}^{\infty}\frac{dl^{2}}{l^{2}}\,\mathcal{F}(x,l^{2})\right)^{2}
+\mathcal{F}(x,k^{2})
\int_{k^{2}}^{\infty}\frac{dl^{2}}{l^{2}}
\ln\!\left(\frac{l^{2}}{k^{2}}\right)\mathcal{F}(x,l^{2})
\Bigg\}.
\end{aligned}
\end{equation}

For generalizations of the equation to take into account DGLAP terms and kinematical constraint corrections \cite{Kwiecinski:1997ee,Kutak:2003bd,Kutak:2025wjy} it is convenient to rewrite the equation as a double integral equation:
\begin{equation}
\label{eq:faneq1}
\begin{aligned}
\mathcal{F}(x,k^{2})
={}&\;\mathcal{F}_{0}(x,k^{2})
+\bar{\alpha}_{s}\int_{x/x_{0}}^{1}\frac{dz}{z}
\int_{0}^{\infty}\frac{dl^{2}}{l^{2}}\,
\Bigg[
\frac{\,l^{2}\mathcal{F}\!\left(\tfrac{x}{z},l^{2}\right)
      -k^{2}\mathcal{F}\!\left(\tfrac{x}{z},k^{2}\right)\,}
     {\lvert k^{2}-l^{2}\rvert}
+\frac{k^{2}\mathcal{F}\!\left(\tfrac{x}{z},k^{2}\right)}
      {\sqrt{4l^{4}+k^{4}}}
\Bigg]
\\[0.8ex]
&\;-\frac{2\bar\alpha_{s}^{2}\pi^2}{N_{c}^2R_p^{2}}
\int_{x/x_{0}}^{1}\frac{dz}{z}\,
\Bigg\{
\Bigg[\int_{k^{2}}^{\infty}\frac{dl^{2}}{l^{2}}\,
\mathcal{F}\!\left(\tfrac{x}{z},l^{2}\right)\Bigg]^{2}
+\mathcal{F}\!\left(\tfrac{x}{z},k^{2}\right)
\int_{k^{2}}^{\infty}\frac{dl^{2}}{l^{2}}\,
\ln\!\Bigg(\frac{l^{2}}{k^{2}}\Bigg)\,
\mathcal{F}\!\left(\tfrac{x}{z},l^{2}\right)
\Bigg\}.
\end{aligned}
\end{equation}
where ${\cal F}_0(x,k^2)$ is a starting distribution.
Another generalization is to account for nuclear target. This ammounts to replacing ${\cal F}(x,k^2)$ by  $A\,\,{\cal F}(x,k^2)$, where $A$ is a mass number, on both sides of eq. (\ref{eq:faneq1}) substituting $R_A=R_p A^{1/3}$ and normalizing by mass number.
We then get 

\begin{equation}
\label{eq:faneq2}
\begin{aligned}
\mathcal{F}_{HI/A}\!\left(x,k^{2}\right)
={}&\;\mathcal{F}_{0\,HI/A}\!\left(x,k^{2}\right)
+\bar{\alpha}_{s}\!\int_{x/x_{0}}^{1}\frac{dz}{z}
\int_{0}^{\infty}\frac{dl^{2}}{l^{2}}\,
\Bigg[
\frac{l^{2}\,\mathcal{F}_{HI/A}\!\left(\tfrac{x}{z},l^{2}\right)
      -k^{2}\,\mathcal{F}_{HI/A}\!\left(\tfrac{x}{z},k^{2}\right)}
     {\lvert k^{2}-l^{2}\rvert}
+\frac{k^{2}\,\mathcal{F}_{HI/A}\!\left(\tfrac{x}{z},k^{2}\right)}
      {\sqrt{4l^{4}+k^{4}}}
\Bigg]
\\[0.6ex]
&\;-\frac{2\,A^{1/3}\,\bar{\alpha}_{s}^{2}\,\pi^{2}}{N_{c}^{2}\,R_p^{2}}
\int_{x/x_{0}}^{1}\frac{dz}{z}\,
\Bigg\{
\Bigg[\int_{k^{2}}^{\infty}\frac{dl^{2}}{l^{2}}\,
\mathcal{F}_{HI/A}\!\left(\tfrac{x}{z},l^{2}\right)\Bigg]^{2}
\\[-0.2ex]
&\hspace{6.6em}
+\mathcal{F}_{HI/A}\!\left(\tfrac{x}{z},k^{2}\right)
\int_{k^{2}}^{\infty}\frac{dl^{2}}{l^{2}}\,
\ln\!\left(\frac{l^{2}}{k^{2}}\right)\,
\mathcal{F}_{HI/A}\!\left(\tfrac{x}{z},l^{2}\right)
\Bigg\}.
\end{aligned}
\end{equation}
\begin{figure}[!t]
    \captionsetup{font={small,it}}  
    \centering
    \begin{minipage}{0.65\textwidth}
        \centering
        \includegraphics[width=\textwidth]{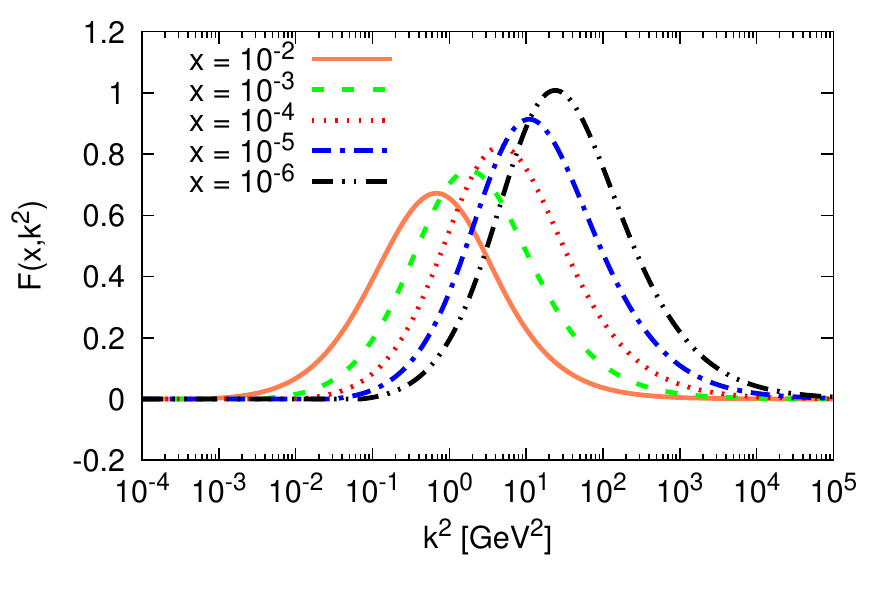}
        \caption{Solution of BK equation from  as  function of $k^2$ for various values of $x$. The figure is taken from \cite{Kutak:2025wjy}.}
        \label{fig:BKrc_k2}
    \end{minipage}
\end{figure}
Once the BK equation is written for dipole gluon density one can directly see why as $k^2$ gets smaller the gluon density gets suppressed for small $k^2$ giving rise to distribution which vanishes at  small values of transverse momenta see fig. (\ref{fig:BKrc_k2}). Namely the integral over the nonlinear term is performed over longer domain what suppresses more and the linear part of the equation as transverse momentum become smaller. Furthermore, one can see that for infinitely large proton i.e. $R_p\rightarrow\infty$ the nonlinear corrections do not matter. One can also see in eq. (\ref{eq:faneq2}) that for heavier nucleus the saturation corrections are larger as the nonlinear term gets multiplied by $A^{1/3}$.\\
The eq. (\ref{eq:faneq1}) can be  obtained directly in the momentum space \cite{Kutak:2006rn}. The derivation is based on considering propagation of regeizzed gluons in the channel taking into account Reggezation and $2\to2$ and $2\to 4$ transition kernels projected onto color singlets. 

\section{Resummed BK}
In this section I present one more representation of the eq. (\ref{eq:faneq1}). The BK equation can be written using Regge form factor that resummes unresolved and virtual emissions. The transformation leading the equation below relies on performing Mellin transform w.r.t $x$ and combining unresolved and virtual contributions and performing inverse Mellin transform.
For detailed derivation see \cite{Kutak:2012qk}\footnote{The resummed form has been also obtained for eq. (\ref{eq:wwglue}) \cite{Kutak:2011fu} }. The unresolved emissions are defined as those which obey condition $\mu^2>q^2$ where $\mu^2$ is a scale which divides the momenta into resolved and unresolved emissions and $q^2$ is square of transverse momentum emitted in the t-channel see fig. (\ref{fig:plot1-tpv}). The resummed equation reads:

\begin{equation}
\label{eq:nonlinear-evolution}
\begin{aligned}
\mathcal{F}\!\left(x,k^{2}\right)
&=
\widetilde{\mathcal{F}}_{0}\!\left(x,k^{2}\right)
+\bar\alpha_{s}\int_{x/x_{0}}^{1}\frac{dz}{z}\,
\Delta_{R}(z,k,\mu)\,
\Bigg\{
\int \frac{d^{2}\bm{q}}{\pi\,q^{2}}\,
\Theta\!\left(q^{2}-\mu^{2}\right)\,
\mathcal{F}\!\left(\frac{x}{z},\,|\bm{k}+\bm{q}|^{2}\right)
\\[0.8ex]
&\hspace{5.0em}
-\frac{2\bar\alpha_{s}^{2}\,\pi^2}{N_{c}^2R_p^{2}}
\Bigg[
\left(\int_{k^{2}}^{\infty}\frac{dl^{2}}{l^{2}}\,
\mathcal{F}\!\left(\tfrac{x}{z},l^{2}\right)\right)^{2}
+\mathcal{F}\!\left(\tfrac{x}{z},k^{2}\right)
\int_{k^{2}}^{\infty}\frac{dl^{2}}{l^{2}}\,
\ln\!\left(\frac{l^{2}}{k^{2}}\right)\,
\mathcal{F}\!\left(\tfrac{x}{z},l^{2}\right)
\Bigg]
\Bigg\}.
\end{aligned}
\end{equation}
with resummed initial distribution expressed in terms of unresummed initial distribution ${\cal F}_0(k,k^2)$ 
\begin{equation}
\widetilde{\mathcal{F}}_{0}\!\left(x,k^{2}\right)\equiv
\frac{1}{2\pi i}\int_{c-i\infty}^{c+i\infty} d\omega\;
x^{-\omega}\,\widehat{\mathcal{F}}_{0}\!\left(\omega,k^{2}\right).
\end{equation}
where 
\begin{equation}
\widehat{\mathcal{F}}_0\!\left(\omega,k^{2}\right)
=
\int_{0}^{x_{0}} dx\; x^{\omega-1}\,\mathcal{F}_0\!\left(x,k^{2}\right).
\end{equation}
and Regge form factor
\begin{equation}
\Delta_{R}(z,k,\mu)\equiv
\exp\!\left[
-\bar\alpha_{s}\,
\ln\!\left(\frac{1}{z}\right)\,
\ln\!\left(\frac{k^{2}}{\mu^{2}}\right)
\right].
\end{equation}

\section*{Summary}
In this paper, I presented the BK equation for the dipole gluon density for both a proton and a nucleus. In particular, I provided all intermediate steps required to go from Eq.~(\ref{eq:wwglue}) to Eq.~(\ref{eq:faneq1}). I also derived the unfolded form of the nonlinear term, in which the structure of the triple-Pomeron vertex becomes explicit. Furthermore, I discussed the impact-parameter dependence, which leads to a form of the BK equation in which the proton radius appears explicitly in the nonlinear term. In the final part of the note, I discussed the resummed version of the BK equation.
\section*{Acknowledgements}
I would like to thank K. Golec-Biernat, W. Li, A. Stasto, R. Straka, F. Salazar, P. Jacobos, P. Caucal, R. Bussarie, P. Kotko for discussions on the various aspects of the BK equation for dipole gluon density.

\bibliographystyle{JHEP} 
\bibliography{references}

\end{document}